\documentclass{aa}
\usepackage{amsmath,mathrsfs}
\usepackage[table,xcdraw]{xcolor}
\usepackage{lscape}
\usepackage{capt-of}
\usepackage{graphicx}
\usepackage{placeins}
\usepackage[utf8]{inputenc}
\usepackage{filecontents}
\usepackage{txfonts}
\usepackage{natbib}
\usepackage{multicol,multirow}
\newcommand{\Gaia}{{\it Gaia}}

\newcommand{\Gbp}{$G_{\rm BP}$}
\newcommand{\Grp}{$G_{\rm RP}$}
\newcommand{\Teff}{$T_{\rm eff}$}
\newcommand{\logg}{$log(g)$}
\newcommand{\xgb}{\texttt{XGBoost}}
\newcommand{\feh}{${\rm [Fe/H]}$}
\newcommand{\kms}{km~s$^{-1}$}
\newcommand{\vlos}{$v_{\rm los}$}
\newcommand{\Lagr}{\mathscr{L}}

\begin{document}

\title{Leaves on trees: identifying halo stars with extreme gradient boosted trees}

\titlerunning{Identifying halo stars with extreme gradient boosted trees}

\author{Jovan Veljanoski
        \and Amina Helmi
        \and Maarten Breddels
        \and Lorenzo Posti}

   \institute{Kapteyn Astronomical Institute, University of Groningen,
              Landleven 12, 9747 AD Groningen, The Netherlands\\
              \email{jovan@astro.rug.nl}}

  \date{\today}


  \abstract
  {Extended stellar haloes are a natural by-product of the hierarchical
  formation of massive galaxies like the Milky Way. If merging is a
  non-negligible factor in the growth of our Galaxy, evidence of such events
  should be encoded in its stellar halo. Reliable identification of genuine halo
  stars is a challenging task however.}
  {With the advent of the \Gaia\, space telescope, we are ushered into a new era
  of Galactic astronomy. The 1st \Gaia\, data release contains the positions,
  parallaxes and proper motions for over 2~million stars, mostly in the Solar
  neighbourhood. The 2nd \Gaia\, data release will enlarge this sample to over
  1.5~billion stars, the brightest $\sim 5$~million of which will have a
  full phase-space information. Our aim for this paper is to develop a machine
  learning model for reliably identifying halo stars, even when their full
  phase-space information is not available.}
  {We use the Gradient Boosted Trees algorithm to build a supervised halo star
  classifier. The classifier is trained on a sample stars extracted from the
  \Gaia\, Universe Model Snapshot, which is also convolved with the errors of
  the public TGAS data, which is a subset of \Gaia~DR1, as well as with the
  expected uncertainties for the upcoming \Gaia~DR2 catalogue. We also trained
  our classifier on a dataset resulting from the cross-match between the TGAS
  and RAVE catalogues, where the halo stars are labelled in an entirely model
  independent way. We then use this model to identify halo stars in TGAS.}
  {When full phase-space information is available and for \Gaia~DR2-like
  uncertainties, our classifier is able to recover 90\% of the halo
  stars with at most 30\% distance errors, in a completely unseen test
  set, and with negligible levels of contamination. When line-of-sight
  velocity is not available, we recover $\sim 60\%$ of such halo
  stars, with less than 10\% contamination.  When applied to the TGAS
  catalogue, our classifier detects 337 high confidence red giant
  branch halo stars. At first glance this number may seem small, however it is
  consistent with the expectation from the models, given the uncertainties in
  the data. The large parallax errors are in fact the biggest limitation in our
  ability to identify a large number of halo stars in all the cases studied.}
  {}


   \maketitle
%

\section{Introduction}

Our current understanding of galaxy assembly is that it occurs in a hierarchical
manner: smaller dark matter haloes merge to form larger, more massive objects.
This process results in the formation of diffuse stellar haloes which envelop
massive galaxies like our Milky Way \citep[e.g.][]{2011ApJ...733L...7H,
2010MNRAS.406..744C}. Even though they comprise less than 1\% of the total
stellar mass of a galaxy, in principle stellar haloes keep records of most past
merger and accretion events their host has experienced, encoded in their
dynamics, chemistry, age, spatial structure and star-formation history of their
stellar populations. Thus, detailed studies of the stars comprising galactic
haloes enable us to unravel the formation and evolution history of massive
galaxies, and with that to test the $\Lambda$CDM paradigm.

Due to the close proximity, it is only natural that the most detailed studies of
stellar haloes have been made in the Local Group. Deep, wide-field ground based
photometric surveys such as SDSS and PanSTARRS have been able to uncover a
variety of stellar streams in the Milky Way \citep[e.g.][]{2006ApJ...642L.137B,
2006ApJ...643L..17G,2006ApJ...645L..37G,2002ApJ...569..245N,2014ApJ...787...19M,
2014MNRAS.443L..84B,2016MNRAS.463.1759B,2016ApJ...820...58B}. These are thought
to be the remnants of tidally disrupted dwarf galaxies or star clusters, and
thus can serve as markers of recent accretion events. Indeed, we currently
witness how the Sagittarius dwarf galaxy is being accreted onto the Milky Way
\citep[e.g.][]{1994Natur.370..194I,1995MNRAS.277..781I,2012ApJ...750...80K,
2013ApJ...762....6S}, which serves as evidence that the Galactic halo is still
actively evolving. The Milky Way is not unique in this respect. Stellar streams
are also found in our closest massive neighbour, M31
\citep[e.g.][]{2009Natur.461...66M,2014ApJ...780..128I}, as well as in several
more distant galaxies \citep[e.g.][]{2008ApJ...689..184M,2010AJ....140..962M,
2016ApJ...823...19C}, typically in their outer haloes.

In the Milky Way, the spatially coherent stellar streams are often used to
constrain the enclosed mass and the shape of the underlying gravitational
potential \citep[e.g.][]{2009ApJ...703L..67L,2010ApJ...718.1128L,
2013MNRAS.433.1826S,2016ApJ...833...31B}. They do not paint the full picture,
however. A significant fraction of the stars in the halo of our Galaxy also
belongs to the so called smooth component. This component may be a remnant of
the initial stages of the Milky Way's formation, and may contain information
about the very first objects that built our galaxy. These primordial building
blocks have likely sunk deeper in the Galactic potential well, and due to the
shorter dynamical time-scales are now likely to be fully phased mixed and can
not be readily observed on the sky as coherent structures
\citep{1999MNRAS.307..495H}. In order to detect them, one needs to know
accurately the 3D positions and velocities of their constituent stars
\citep[e.g.][]{2000MNRAS.319..657H}.

At least for now, the Milky Way is the only massive galaxy for which we can
obtain accurate enough astrophysical measurements of its stars hoping to
decipher its assembly history. With the advent of the \Gaia\, satellite we enter
a new era of Galactic astrophysics. The primary data set of \Gaia\, Data Release
1 (DR1) provides the positions, parallaxes and mean proper motion for over 2
million stars in common with the \emph{Tycho-2} and the \textsc{hipparcos}
catalogues, otherwise known as the \emph{Tycho-Gaia} Astrometric Solution
(TGAS). The secondary dataset of \Gaia~DR1 consists of on-sky positions for
over 1~billion sources and their mean $G$-band magnitudes
\citep{2016A&A...595A...2G}. The biggest step forward will happen with the
second data release (DR2), scheduled for the end of April 2018, which will contain 3D
positions, proper motions and optical photometry ($G$, \Gbp, \Grp) for over
1~billion stars. For the brightest $\sim 3-5$~million of those, full phase-space
information will be available. This will enable us to dig deeper in our galaxy's
history than ever before.

Before being able to explore the Milky Way's past, we need a method of
identifying halo stars. In the pre-\Gaia\, era, candidate halo stars were
selected based on their proper motions and metallicities
\citep[e.g.][]{1996AJ....112..668C,1992ApJS...78...87M,1996ApJ...459L..73M,
2009MNRAS.399.1223S}. Such samples may be incomplete since they are biased
against stars having small proper motions. More recently, as surveys became
wider and deeper, studies have also focused on very distant stars in order to
constrain the properties of the halo outside the Solar neighbourhood. Commonly
used halo tracers are RR~Lyrae \citep[e.g][]{2009MNRAS.398.1757W,
2010ApJ...708..717S,2013ApJ...763...32D,2013ApJ...765..154D,2015MNRAS.446.2251T}
which can be pre-selected based on their colours and magnitudes, and then their
distance can be determined via their period-luminosity relation.
Other commonly used halo probes are blue horizontal branch (BHB) stars
\citep[e.g.][]{2008ApJ...684.1143X,2011ApJ...738...79X,2012MNRAS.425.2840D,
2017MNRAS.470.1259D} which can be selected through colour
and magnitude cuts and can act as standard candles. While not exactly standard
candles themselves, main-sequence turn off (MSTO) stars can also provide a
distance range estimate and have been used as halo tracers
\citep[e.g.][]{2008ApJ...680..295B,2010AJ....140.1850B,2017arXiv170602301K}.
K-Giant stars are also a typical way to map the Milky Way halo
\citep[e.g.][]{1980ApJS...44..517B,1990AJ....100.1191M,2000AJ....119.2254M,
2009ApJ...698..567S,2014ApJ...784..170X,2017MNRAS.470.1259D}. Although
not standard candles, their distance can be determined (with approximately 20\%
uncertainty) from measurements of their astrophysical parameters (surface
gravity, metallicity and temperature) obtained from their spectra. These stars
are particularly valuable kinematic tracers of the Milky Way's halo due to their
high intrinsic luminosity.

Recently, \citet{2017A&A...598A..58H} combined \Gaia~DR1 data with the ground
based spectroscopic survey RAVE \citep[DR5][]{2017AJ....153...75K}, and selected
a local sample of halo stars based on their metallicity, which was further
cleaned from disk contaminants by simple kinematic fits. These authors were then
able to use that sample to constrain the overall degree of substructure of the
halo in the Solar neighbourhood. Inspired by their work, and the upcoming
\Gaia\, Data Release 2 (DR2), in this paper we present a supervised method for
selecting halo stars. The method can be entirely data driven, and uses the
position, velocity, photometry and if available the metallicity of the stars to
make the classification. This exploits the fact that halo stars have distinctly
different kinematics compared to disk stars. In addition, it is expected that
the stellar halo is more metal-poor than the other Galactic components
\citep[e.g.][]{1978ApJ...225..357S,2001ApJ...554.1044C}, and thus its
constituent stars are expected to have bluer colours on a HR
diagram, which further improves the selection. We determine how viable our
method is for selecting halo stars within the Solar neighbourhood with \Gaia\,
DR2 data when having 5D phase space information only (without line-of-sight
velocities) and \Gaia\, \Gbp, \Grp, and $G$ optical magnitudes.

This paper is structured as follows: In Section~\ref{sec:method} we introduce
the supervised classifier used for identifying halo stars. In
Section~\ref{sec:gums} we test the viability of this algorithm to reliably
select halo stars using the Gaia Universe Model Snapshot and  mock catalogues
resembling the currently available TGAS and the upcoming \Gaia~DR2 datasets. We
then apply our model to the TGAS subset of \Gaia~DR1 data in
Section~\ref{sec:tgas} and present our conclusions in Section~\ref{sec:concl}.

\section{Methodology}
\label{sec:method}

\subsection{Gradient Boosted Trees}
\label{subsec:xgb}

To build a model for classifying halo stars we use a technique called Gradient
Boosted Trees \citep{2001Ann.Stat.29.5.1189F}. In machine learning, boosting is
an ensemble technique where new models are added in order to improve on the
errors made by previous models. The models are added sequentially until no
further improvement is made. Gradient boosting is an approach where the new
models are created based on the residuals of prior models, which then are added
together to make a final prediction. The term ``gradient boosting'' comes from
the usage of the gradient descent algorithm which is used to minimize an
arbitrary differentiable loss function when adding new models. The models in
this case are decision trees.

Gradient Boosted Trees is a powerful modelling technique with high predictive
power. By combining many simple decision tree models, one can describe
complicated and non-linear relationships amongst different features in a
dataset. Since the base model is a decision tree, the final combined model can
be easily interpreted. This is significantly better than some of the
competing machine learning algorithms such as Support Vector Machines and
Artificial Neural Networks, where it is difficult to understand how the
classification boundary was drawn, especially when dealing with datasets having
high dimensionality. When using Boosted Trees, in contrast to most other
algorithms, one does not need to scale the data or do feature engineering. In
addition, this technique is robust to uninformative features, meaning that there
is no penalty when training the model while using features that do not add any
information towards the classification objective. This method also allows the
usage of sparse data. This can be quite useful, since \Gaia~DR2 may contain
\Teff\, for a subset of stars, which may add information to the classification
process.

Historically, the major drawback of Gradient Boosted Trees is the long duration
of the training process. The need to build many decision trees in succession
makes this procedure difficult to parallelize and optimize, making it hard to
scale up to datasets having a large number of samples or features. In this
paper we use \xgb\footnote{\url{https://github.com/dmlc/xgboost}}
\citep{2016arXiv160302754C}, a state-of-the-art open source implementation of
the Gradient Boosted Trees technique which mitigates the performance
issues typically associated with this technique. It employs a novel tree
building algorithm based on the studies by \citet{Li:2007:MLR:2981562.2981675,
Bekkerman:2011:SUM:2107736.2107740,Tyree:2011:PBR:1963405.1963461} and is able
to use all available CPUs in a machine during training. \xgb\, also supports
distributed training, so one could use a computer cluster for building up a
model, as well as out-of-core computing for datasets that are too large to fit
into memory. This makes \xgb\, a suitable tool for creating models using large
tabular datasets such as \Gaia~DR2 and its subsequent releases.

\subsection{Training the classifier}
\label{subsec:training}

To train the \xgb\, classifier, we first select the input data and associated
features. These will be described in more detail in Sections~\ref{sec:gums} and
\ref{sec:tgas}. The data is split into a training and testing sets having
proportions of $70-30\%$ respectively. The split is stratified, meaning that it
preserves the fraction of halo to non-halo stars in both the training and
testing sets.

\xgb\, features various hyperparameters that define the behaviour of the
classifier model. These parameters set the number of decision trees in the
ensemble and their maximum depth, the number of features to consider when
growing a tree, and control the process through which a tree is grown and
pruned. One can create an optimal model for a given dataset by properly tuning
these hyperparameters.

We use the Bayesian optimization package
\texttt{bayes\_opt}\footnote{\url{https://github.com/fmfn/BayesianOptimization}}
based on \citet{2010arXiv1012.2599B,2012arXiv1206.2944S} to find the
hyperparameters values that minimize the loss function of the \xgb\, classifier.
Bayesian optimization works by constructing a posterior distribution function
in which every variable is normally distributed, and every collection of random
variables can be described by a multivariate Gaussian distribution, otherwise
known as a Gaussian process. For each iteration, such a function is fitted to
the known samples of the loss function, and then an exploration algorithm is
employed to determine the combination of parameters to be used for sampling the
loss function in the next step. This is an ideal method of optimizing a function
for which the sampling is very computationally expensive, since it minimizes the
number of iterations needed to find a parameter combination that is close to
optimal.

At each step of the Bayesian optimization, the value of the loss function is
taken to be the mean loss coming from a 5-fold cross-validation. This is done
as follows. First, the training set is split into 5 subsets in a stratified
manner. The model is then trained on 4 of those subsets, while the remaining
subset is used to asses the model performance based on the Logarithmic Loss
function. In machine learning, the logarithmic loss ($\Lagr$) quantifies the
accuracy of a classifier by penalizing false classifications, and in the case of
only two classes it is defined as:

\begin{equation}
\Lagr =
-\frac{1}{N}\sum_{i=1}^{N}[y_i\, {\rm log}(p_i) + (1-y_i)\, {\rm log}(1-p_i)]
\label{eq:logloss}
\end{equation}
where $N$ is the number of samples, $y_i$ is a binary indicator of whether a
sample has been correctly identified and $p_i$ is the probability of assigning a
positive label to that sample. This is repeated five times, and each time a
different of the five subsets is used to evaluate the model. Throughout this
process we also monitor the precision, recall, and the Matthews correlation
coefficient. In the following definitions of these metrics $TP$ stands for
``true positives'', i.e. stars that were correctly labelled as halo, $TN$ stands
for ``true negatives'' or stars that were correctly identified as non-halo,
$FP$ or ``false positives'' are stars wrongly assigned as halo and $FN$ or
``false negatives'' are halo stars what have been mislabelled as non-halo.
With this in mind, the recall can be expressed as:

\begin{equation}
recall = \frac{TP}{TP+FN}
\label{eq:recall}
\end{equation}
and measures the fraction of the positive class that is being recovered, and in
this case that is the fraction of correctly identified halo stars. A value of
1 means that all halo stars were successfully recovered. The precision is
defined as:
\begin{equation}
precision = \frac{TP}{TP+FP}
\label{eq:precision}
\end{equation}
which measures the level of contamination present in the sample of positive
predictions. A precision value of 0.8 means that there is 20\% contamination
in the predicted sample of halo stars.

The Matthews correlation coefficient is designed to measure the performance of
a binary classifier and it takes into account both the false positive and false
negative predictions, and it is defined as:

\begin{equation}
MCC = \frac{TP \times TN - FP \times FN}{\sqrt{(TP+FP)(TP+FN)(TN+FP)(TN+FN)}}
\label{eq:mcc}
\end{equation}
A value of 1 signifies a perfect model, 0 represents random guessing while a
value of -1 means a total disagreement between the features and the
classification.

Monitoring these metrics simultaneously is very important when trying to asses
a model which is designed to predict the minority class of a highly imbalanced
dataset. For the final several steps of the Bayesian optimization procedure, we
check that the standard deviations of the logarithmic loss, precision, recall
and in the Matthews correlation coefficient are small. The small deviation of
these statistics tells us that the model performance is stable when the training
and the testing data is varied, meaning that it does not suffer from
over-fitting. In general we find the \xgb\, classifier to perform quite well for
a sensible choice of parameters, and we observe only a slight improvement in the
results after undergoing the rigorous tuning described above.

Once the hyperparameters of the classifier are tuned and we are satisfied with
the performance of the model judging from the the cross-validation scores
(recall, precision, Matthews correlation coefficient) we proceed to evaluate
the predictive power of the model via the so far unused test set.
In Section~\ref{sec:gums} we present and discuss the results
obtained from this final evaluation of the model.

\section{Identifying halo stars in GUMS}
\label{sec:gums}

To test the viability of our classifier to detect halo stars, we first apply it
to data extracted from the \Gaia\, Universe Model Snapshot
\citep[GUMS][]{2012A&A...543A.100R}, based on the Besan\c{c}on Galaxy Model
\citep{2003A&A...409..523R}. Due to how GUMS is generated, the stellar sources
are labelled to belong to one of the four galactic components, thin and thick
disk, bulge and halo. Here we focus on how well we can identify halo stars
TGAS-like and \Gaia~DR2-like data. To create a TGAS-like set, we produce a
magnitude limited sample that is used to evaluate the best possible performance
of the model in the Solar neighbourhood. This is an ideal scenario because the
data is perfect as it contains no measurement or systematic uncertainties. We
then convolve this sample with the median uncertainties from the TGAS data, in
order to evaluate the performance of our model under more realistic conditions.
We also select a much larger sample from GUMS, which we error convolve with the
expected uncertainties from \Gaia~DR2.

\subsection{The ideal Solar neighbourhood sample}
\label{subsec:ideal-gums}

The stars in the TGAS-like sample extracted from GUMS have magnitudes between
$6 \leq G \leq 12.5$, but also $0.2 \leq log(g) \leq 5$, $3000 \leq T_{\rm eff}
\leq 9000$~K. The magnitude criterion is chosen such that this sample loosely
resembles the TGAS subset of \Gaia~DR1 \citep{2016A&A...595A...2G,
2016A&A...595A...4L}, while the cuts on \Teff\, and \logg\, are chosen
considering the RAVE~DR5 data. Our TGAS-like sample contains $\sim4.7$ million
stars, of which 9240 (2\%) belong to the halo component according to the labels
in GUMS.

\begin{figure}
\centering
\includegraphics[width=\textwidth/2]{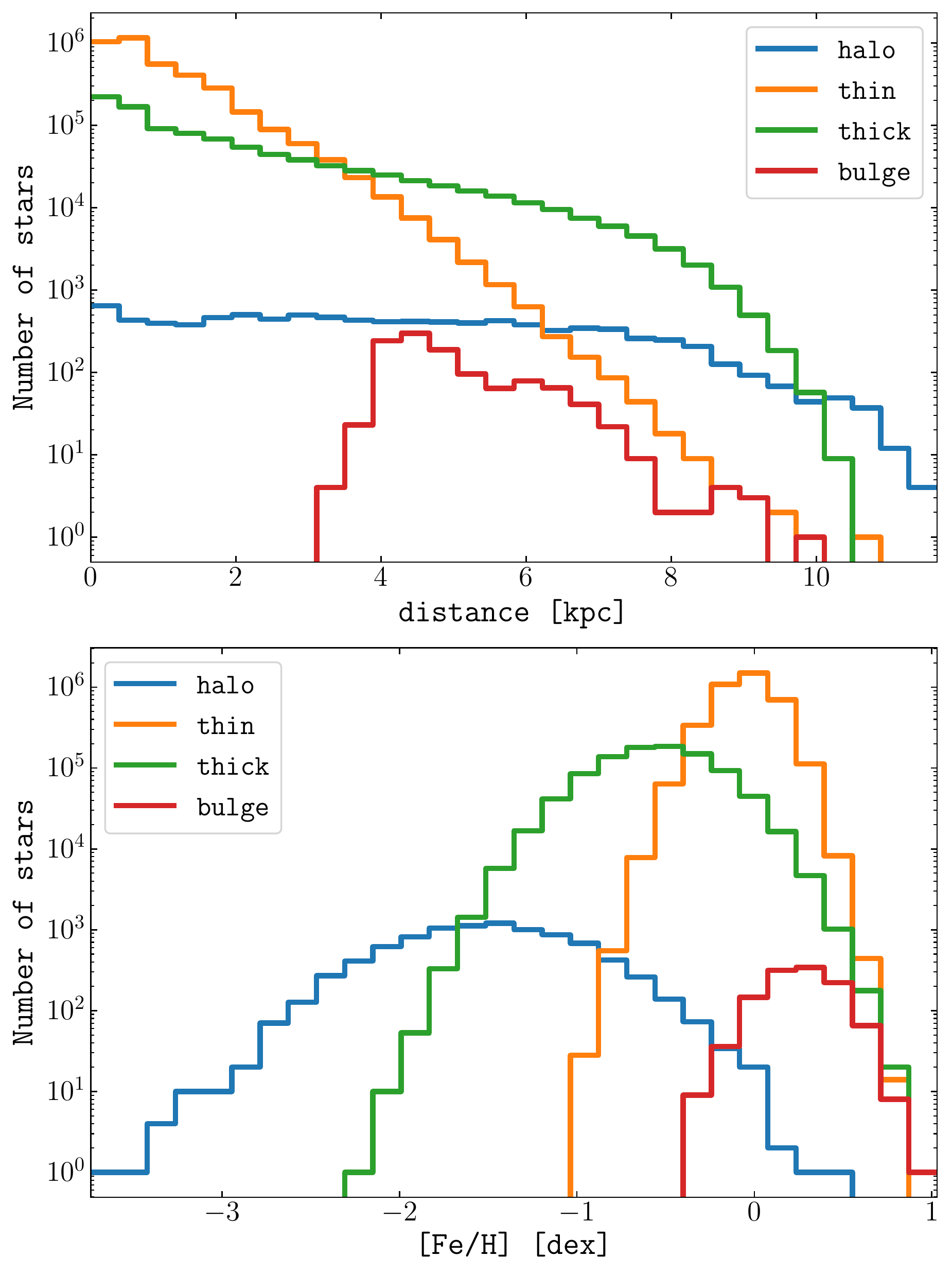}
\caption{Distance (top) and \feh\, (bottom) distributions for each of the
Galactic components of the stars in the Solar Neighbourhood from the
perfect GUMS sample. There are $\sim4784319$ stars in this dataset, 9240 of
which are halo. The halo stars have a roughly constant distribution out to
distance of $\sim 8$~kpc from the Sun. While they span a large range of \feh\,
value, the bottom panel shows that the halo stars are on average more
metal-poor than stars belonging to the other Galactic components.}
\label{fig:gums-dist-feh}
\end{figure}

Figure~\ref{fig:gums-dist-feh} shows the distance and \feh\, distributions for
these stars. From the distance distribution, one can see that the halo number
count is roughly constant out to a distance of $\sim 8$~kpc, and that disk stars
significantly outnumber the nearby ($<2$~kpc) halo. On the other hand, the halo
stars are systematically more metal-poor than the rest, as shown on the bottom
panel in Figure~\ref{fig:gums-dist-feh}.

\begin{figure*}
\centering
\includegraphics[width=6.0cm]{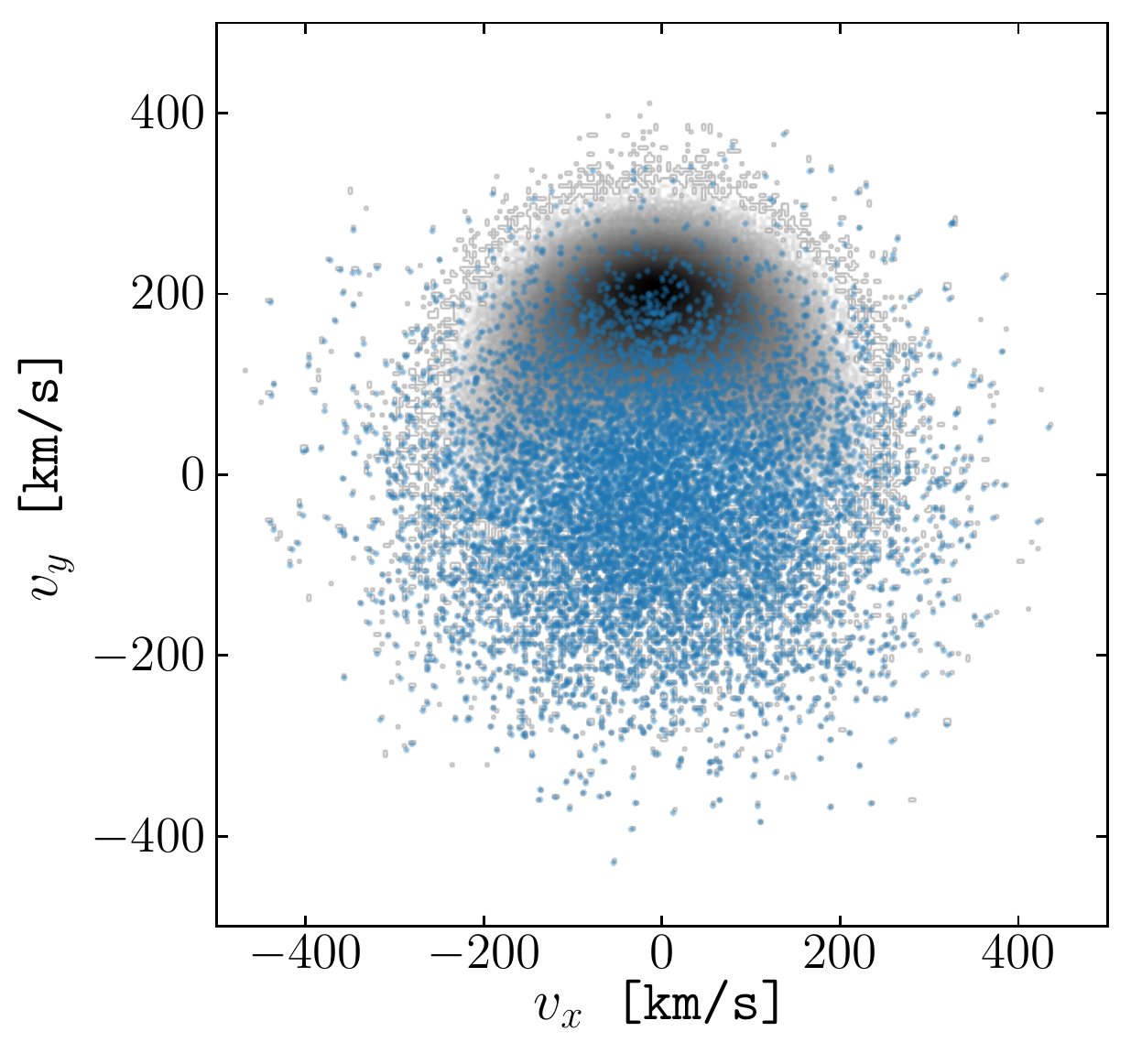}
\includegraphics[width=6.0cm]{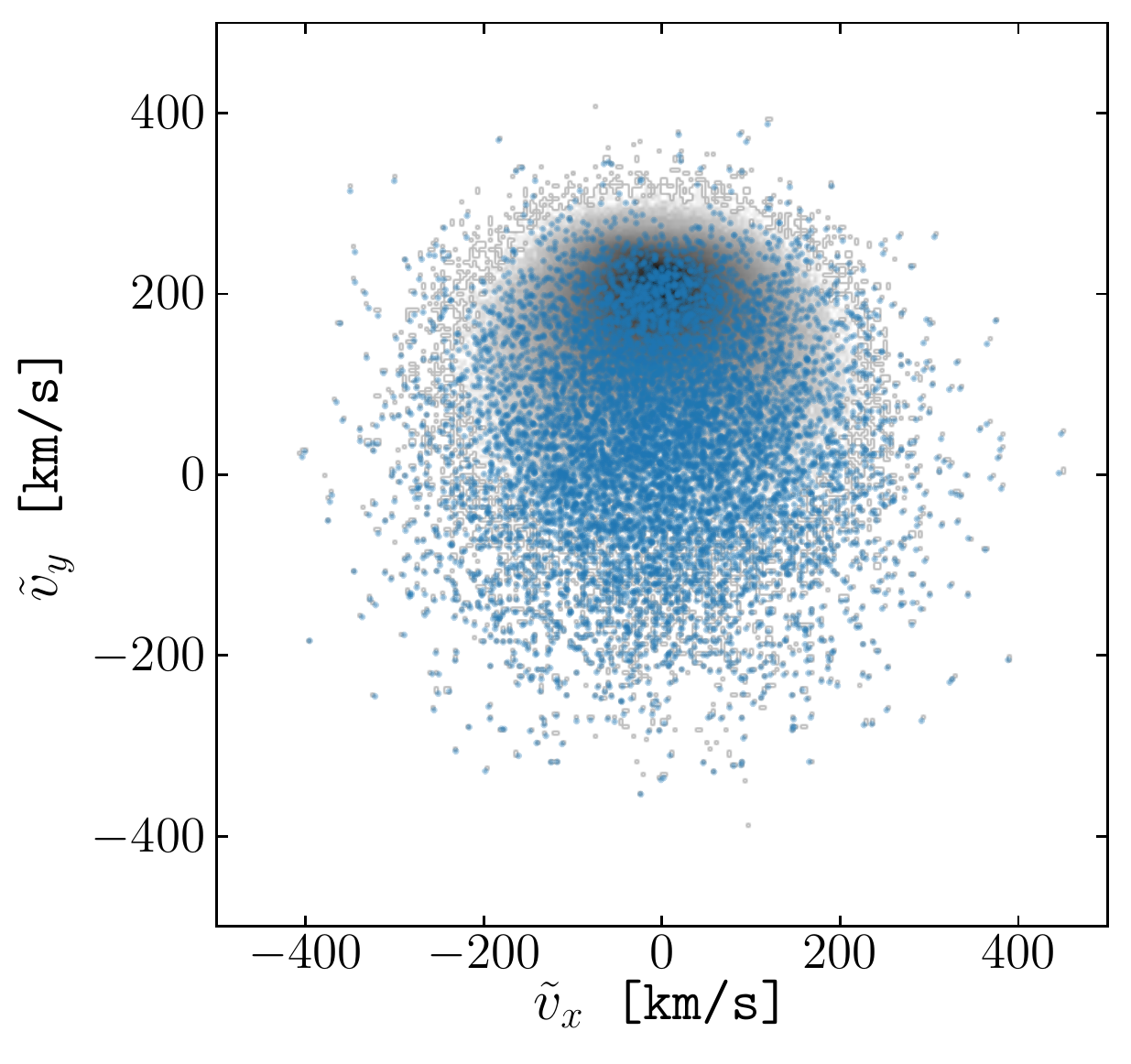}
\includegraphics[width=6.0cm]{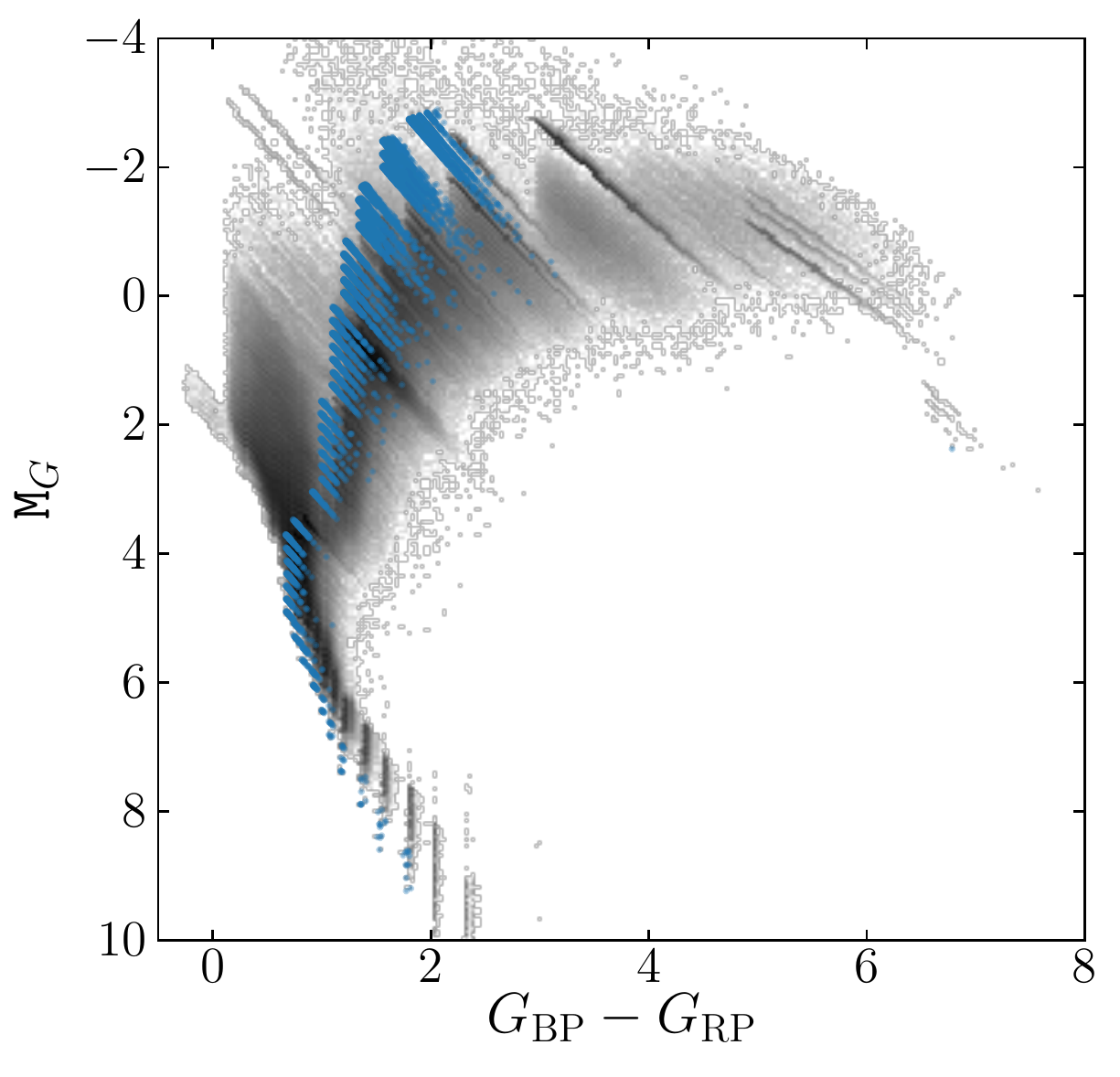}
\caption{The left panel shows the Cartesian velocities of the GUMS sample
in the Solar neighbourhood. The blue points mark the halo stars, while the
underlying density distribution shows all the stars in the sample. One can
readily notice that the halo members have distinctly different kinematics
compared to the rest of the stars. The middle panel shows the equivalent
velocity distribution calculated assuming \vlos\, = 0~\kms, for the case when
the radial velocities of the stars are not known. The halo stars with \vlos\,
close to 0 are not affected by this transformation, while the stars with large
\vlos values cluster around $\tilde{v}_y \sim 200$~\kms. There is still a
kinematic distinction between the halo stars and the rest. The right panel
shows a HR diagram, where the symbols are the same as on the left panel.
The halo stars have systematically bluer colours, in line with their \feh\,
distribution.}.
\label{fig:gums-kin-cmd}
\end{figure*}

On the left panel in Figure~\ref{fig:gums-kin-cmd} we show the velocity
distribution in Cartesian coordinates of our GUMS sample, where the blue points
mark the halo stars. One can see that the halo stars have distinctly different
kinematics than the rest. The disk stars rotate around the Galactic centre with
a mean ${\rm v_y}\sim 200$~\kms, while the stellar halo shows no such orderly
motion, and the mean of its velocity components are centred close to zero. The
right panel on the same Figure shows an HR diagram. The halo stars show
systematically bluer colours compared to the rest, in line with their \feh\,
distribution shown in Figure~\ref{fig:gums-dist-feh}. It is this combination of
their characteristic kinematics, bluer colours and low \feh\, values that we
rely on to train the classifier to separate the halo stars from the rest.

To asses the ability of our classifier to identify halo stars, we consider
three cases depending on the data that may be available for the training
process. In the first case we use the full phase-space information,
together with the absolute $G$ magnitude (M$_G$), the optical \Gbp$-$\Grp\,
colour, and the \feh\, for all stars in our GUMS sets. This is the best case
scenario in terms of data availability. It is unlikely that \feh\, estimates
will be released as part of \Gaia~DR2, however we find it useful to test the
performance of our classifier under optimal conditions.

In the second test case, which is same as above but without metallicity
information, we use features that we know will be available in \Gaia~DR2, at
least down to magnitude $G \sim 12.5$: 3D positions and velocities,
coupled with \Gaia\, photometry. These quantities are expected to be available
for $\sim 5$~million stars in \Gaia~DR2.

Our third scenario is designed to be applicable to the currently available TGAS
dataset: we use the 3D positions, proper motions and \Gaia\, photometry of the
stars, without any knowledge of their line-of-sight velocities. In this case,
instead of transforming the proper motion and line-of-sight velocities
according to the following equations:
\begin{equation}
v_x = v_{\rm los}\,{\rm cos}(l)\,{\rm cos}(b) - k\,d\,\mu_{l}\,{\rm sin}(l) - k\,d\,\mu_{b}\,{\rm cos}(l)\,{\rm sin}(b)
\label{eq:vx}
\end{equation}
\begin{equation}
v_y = v_{\rm los}\,{\rm sin}(l)\,{\rm cos}(b) + k\,d\,\mu_{l}\,{\rm cos}(l) - k\,d\,\mu_{b}\,{\rm sin}(l)\,{\rm sin}(b) + v_{LSR}
\label{eq:vy}
\end{equation}
\begin{equation}
v_z = v_{\rm los}\,{\rm sin}(b) + k\,d\,\mu_{b}\,{\rm cos}(b)
\label{eq:vz}
\end{equation}
we assume \vlos = 0~\kms and derive pseudo-Cartesian coordinates
$(\tilde{v}_x, \tilde{v}_y, \tilde{v}_z)$. In the above equations, $l$
and $b$ are the longitude and latitude of the stars in Galactic
coordinates, while $\mu_l$ and $\mu_b$ are the associated proper
motions. The distance is denoted by $d$ in units of kpc measured from
the Sun, $v_{\rm los}$ are the line-of-sight velocities of the stars
in units of \kms, while $k=4.74057$ is a scaling factor which puts the
velocities ($v_x, v_y, v_z$) in units of \kms. In Eq.~(\ref{eq:vy}) we
correct the $v_y$ velocity for the motion of the Local Standard of
Rest $v_{LSR}$ which we assume to be 220 \kms.  In the middle panel of
Figure~\ref{fig:gums-kin-cmd} we show the distribution of all stars in
our Solar neighbourhood sample in the $\tilde{v}_x-\tilde{v}_y$
space. The halo stars having \vlos~$\sim 0$ ~\kms\, are very weakly
affected by this transformation. This is also true for the majority of
the disk stars in the Solar neighbourhood which are known to have
\vlos $\sim 0$. On the other hand, the halo stars with larger \vlos\,
but small tangential motions are seen to cluster around $\tilde{v}_y
\sim 200$~\kms. In general, even with this transformation, one can see
that there is a clear distinction between the overall kinematics of
the halo stars compared to the vast majority of stars belonging to the
other Galactic components.

\begin{figure}
\centering
\includegraphics[width=8.8cm]{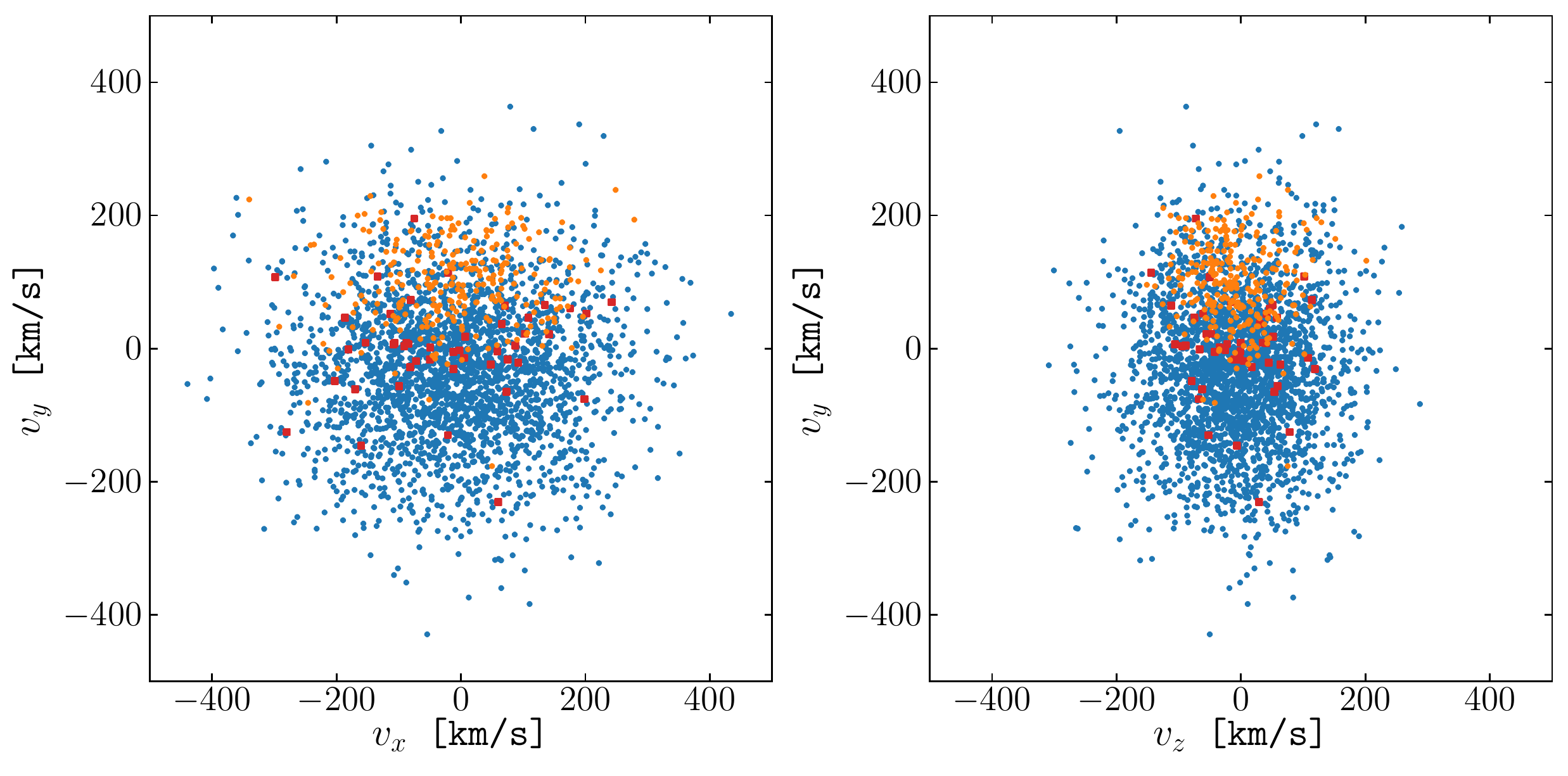}
\caption{Velocity distribution in the $v_x - v_y$ space (left) and in
  the $v_z - v_y$ space (right) of the stars labelled as halo (true
  positives, blue points), stars wrongly labelled as halo (false
  positives, red squares), and stars wrongly labelled as non-halo
  (false negatives, orange points), for the test case when using
  photometry and full phase-space information on the ideal GUMS
  selected sample in the Solar neighbourhood.}
\label{fig:gums-case2-test}
\end{figure}

For each of the three cases described above, we train a Gradient Boosted Trees
classifier as described in Section~\ref{subsec:training}. The results of the
evaluation of these 3 cases are summarized in Table~\ref{tab:GUMS} according
to the metrics we follow during the training and optimization process. From this
Table, one can see that in the cases when we use full phase-space information of
the stars as a training feature, we detect nearly all halo stars in the test
sample. In addition, the level of contamination is negligible. Knowledge of the
metallicity is not necessary, as case 2 shows, especially when there are
no measurement uncertainties the \Gbp$-$\Grp\, colour is an excellent proxy for
\feh\, for the halo stars. Figure~\ref{fig:gums-case2-test} shows the velocity
distribution, corrected for the LSR, for the correctly identified halo stars
(blue), false positives (red) and false negatives (orange). It can be seen that
the stars which were incorrectly labelled as non-halo have kinematics similar to
the disk.

\begin{figure*}
\centering
\includegraphics[width=\textwidth]{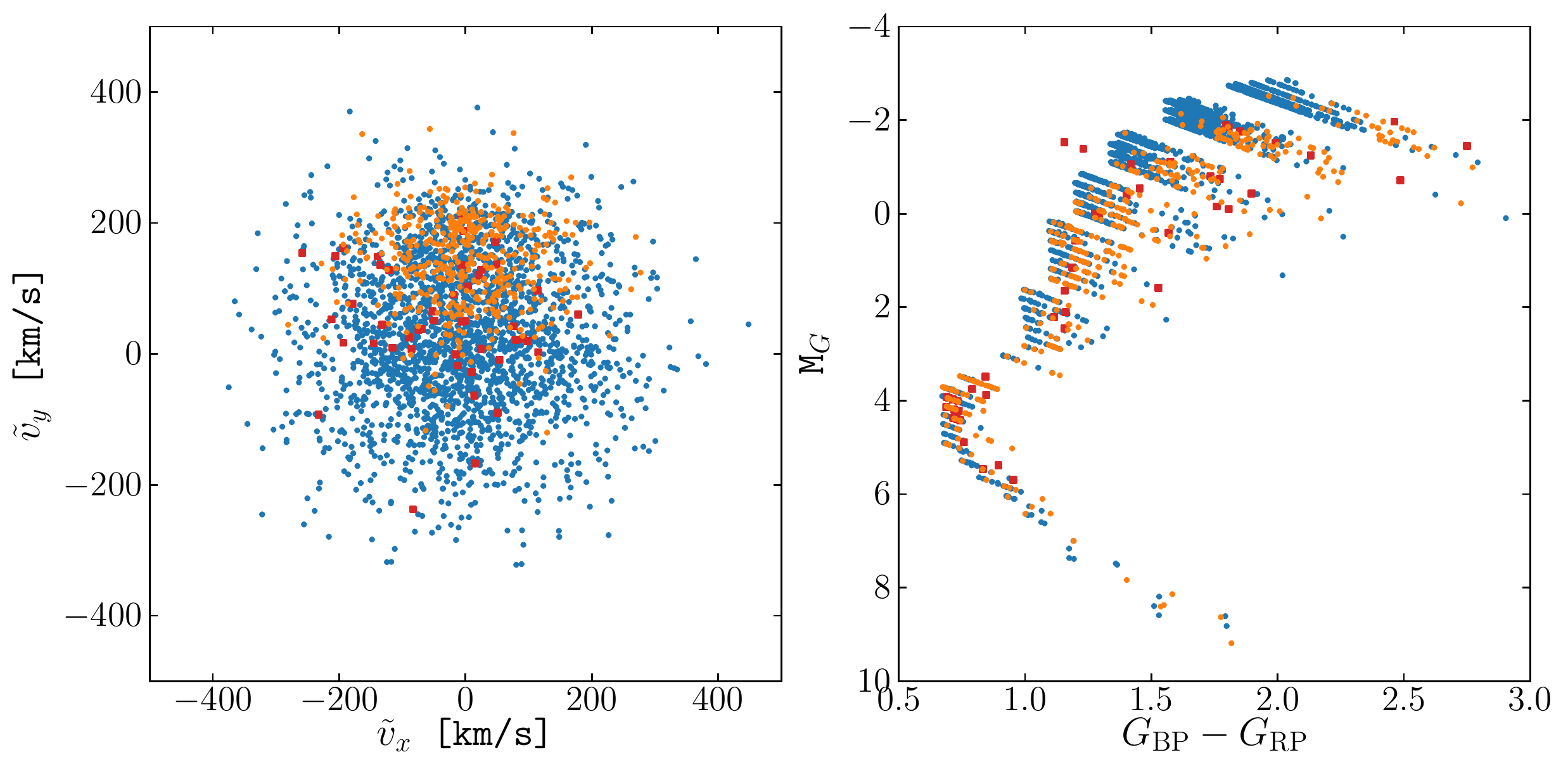}
\caption{Diagnostics for the model trained on the ideal GUMS Solar
neighbourhood sample, using only the \Gaia\, optical photometry and 5D
phase-space information. The left panel displays the velocity distribution in
Cartesian coordinates, assuming \vlos\, = 0~\kms.
The right panel shows a HR diagram. The blue points mark the true positive
(correctly identified halo stars), the orange points mark the false negatives,
while the red squares mark the false positives. Not knowing the line-of-sight
velocities of the test stars makes for increase in the numbers of false negative
detections.}
\label{fig:gums-case3-test}
\end{figure*}

In the final test case, when we do not know the \vlos, we recover over 84\% of
the halo stars. The precision is 98\% which means that the degree of
contamination is negligible. The left panel in
Figure~\ref{fig:gums-case3-test} show the velocity distributions of the
recovered halo stars, as well as that of the false positives, and the halo stars
missed by the classifier (false negatives). It is interesting to see that the
majority of the false negatives have $\tilde{v}_y \sim 200$~\kms, and are in
fact those stars for which the conversion to the Cartesian velocity coordinate
system is least reliable, i.e. for the stars which had larger line-of-sight
velocities and small tangential motions. The right panel on
Figure~\ref{fig:gums-case3-test} shows a HR diagram, where one can
see that the classification is most reliable for stars with very blue
\Gbp$-$\Grp\, colours, especially if they are giants. Moreover, we see no
difference in the metallicity distribution of the correctly classified halo
stars, and stars falsely labelled as non-halo. This mislabelling is mainly due
to not knowing their line-of-sight velocities.

\begin{table*}
\centering
\caption{Metrics describing the performance of our halo star classifier for the
three  different cases depending on the training features available. The upper
bound of the uncertainty of these metrics, calculated as the standard deviation
from the 5-fold cross-validation process during the training, is $0.03$. The
superscripts $a$ and $b$ indicate that the model was evaluated on GUMS
selected samples with $G<12.5$~mag, and $G<19$~mag respectively.}
\label{tab:GUMS}
\begin{tabular}{llll}
\hline
\hline
              & \multicolumn{3}{c}{ideal GUMS sample with $G<12.5$~mag} \\
\hline
                 & 6D phase-space + phot +\feh & 6D phase-space + phot & 5D phase-space + phot \\
\hline
recall           & 0.94                        & 0.90                  & 0.84 \\
precision        & 0.99                        & 0.98                  & 0.98 \\
logarithmic loss & 0.01                        & 0.01                  & 0.01 \\
Matthews coef    & 0.96                        & 0.94                  & 0.90 \\
&&&\\
&&&\\
              & \multicolumn{3}{c}{GUMS sample with $G<12.5$~mag convolved with median TGAS uncertainties} \\
\hline
                 & 6D phase-space + phot +\feh & 6D phase-space + phot & 5D phase-space + phot \\
\hline
recall           & 0.71                        & 0.64                  & 0.46 \\
precision        & 0.90                        & 0.87                  & 0.82 \\
logarithmic loss & 0.01                        & 0.01                  & 0.01 \\
Matthews coef    & 0.80                        & 0.75                  & 0.62 \\
&&&\\
&&&\\
              & \multicolumn{3}{c}{error convolved GUMS sample with \Gaia~DR2 uncertainties} \\
\hline
                 & 6D phase-space + phot +\feh$^{a}$ & 6D phase-space + phot$^{a}$ & 5D phase-space + phot$^{b}$ \\
\hline
recall           & 0.93                              & 0.90                        & 0.58 \\
precision        & 0.98                              & 0.98                        & 0.92 \\
logarithmic loss & 0.01                              & 0.01                        & 0.02 \\
Matthews coef    & 0.95                              & 0.94                        & 0.73 \\
\hline
\end{tabular}
\end{table*}

\subsection{Detecting halo stars in errors-convolved GUMS samples}

The excellent detection rates we report in Section~\ref{subsec:ideal-gums} are
largely due to the dataset being ideal. Since our goal is to identify halo stars
in the TGAS and the upcoming \Gaia~DR2 datasets, we test the performance of our
classifier on two mock catalogues that resemble these datasets.

\subsubsection{TGAS uncertainties}

We create a mock catalogue which resembles TGAS by convolving the observables in
the GUMS sample selected above with the median uncertainties from TGAS listed
in Table~\ref{tab:err}. Since TGAS does not provide \vlos\, and \feh, for these
we took the median uncertainties form the cross-match between TGAS and RAVE.
Since this mock catalogue is not ideal but contains measurement uncertainties,
computing the distance to stars by inverting the associated parallax may not
be reliable in all cases and can lead to biases in the data. Assuming that the
parallax uncertainty distribution is Gaussian, \citet{2017A&A...598A..58H} has
shown that for relative parallax error $\leq 30$\%, the distance computed by
taking the reciprocal values of the parallax is not too biased compared to the
true distance. Therefore we proceed to work only with stars that have relative
parallax error $\leq 30$\%. Such a selection leaves $\sim 3$ million stars in
the GUMS error convolved sample with TGAS uncertainties, of which only 1379 are
halo according to the data model ($\sim 0.046$\% of the sample).

As before, we train the halo star classifier as described in
Section~\ref{subsec:training}. The choice to both train and evaluate the
classifier on the error convolved data makes for a more generalized model.
Training the model on the ideal and applying it on the error convolved set may
cause the classifier to under-perform, since the two sets, training and testing,
would be drawn from different distributions.

The performance of the classifier is shown on Table~\ref{tab:GUMS}. One can see
that after adding TGAS-like uncertainties, the performance of the classifier
does indeed decline. In the worst case scenario, in which we do not have
line-of-sight velocity information the classifier is able to recover only 46\%
of the halo stars, with a contamination level of 18\%. In fact, given the
magnitude of the uncertainties in the observables, the precision of the model
has remained exceptionally high, with the contamination level never reaching
20\%. This means that the halo candidates selected by the classifier have a high
probability of being true halo stars, even after the error convolution.

\subsubsection{\Gaia~DR2 uncertainties}

The primary reason for building this classifier is to detect halo stars in
\Gaia~DR2. To test the ability of the classifier to do this, we create a second
mock catalogue which roughly resembles \Gaia~DR2 by selecting stars in GUMS that
have $G \leq 19$, $0.2 \leq log(g) \leq 5$, and
$3000 \leq T_{\rm eff} \leq 9000$~K. The sample is then convolved with the
expected uncertainties for \Gaia~DR2 according to the following relations
for the astrometry:
\begin{eqnarray}
& \sigma_\pi [{\rm \mu as}]  =  0.9965\, (-1.631 + 680.766\,z_G + 32.732\,z_G^2)^{0.5}\,t_{\rm frac}^{0.5} \nonumber \\
& \sigma_\alpha [{\rm \mu as}]  =  0.787\,\sigma_\pi \nonumber \\
& \sigma_\delta [{\rm \mu as}]  =  0.699\,\sigma_\pi \nonumber  \\
& \sigma_{\mu_\alpha} [{\rm \mu as\, yr^{-1}}]  =  0.556\,\sigma_\pi\,t_{\rm frac} \nonumber \\
& \sigma_{\mu_\delta} [{\rm \mu as\, yr^{-1}}]  =  0.496\,\sigma_\pi\,t_{\rm frac} \nonumber\\
\label{eq:sig_astro}
\end{eqnarray}
where  $z_G = {\rm MAX}[10^{0.4\,(12.09-15)}, 10^{0.4\, (G-15)}]$,
and $t_{\rm frac} = 60/22$ is the total number of months for which the nominal
\Gaia\, mission is scheduled to run over the number of months during which the
DR2 data is observed (A.~G.~A~Brown, private communication).
For the photometry we assume
\begin{eqnarray}
\sigma_G [{\rm mag}] & = & 10^{-3}\, (0.04895\, z_G^2 + 1.8633\,z_G + 0.0001985)^{0.5} \nonumber \\
\sigma_{\rm BP/RP} [{\rm mag}] & = & 10^{-3}\,(10^{a_{BP/RP}}\,z_G^2 + 10^{b_{BP/RP}}\,z_G + 10^{c_{BP/RP}})^{0.5} \nonumber \\
\end{eqnarray}
where $(a_{BP}, b_{BP}, c_{BP})$ = $(1.334, 1.623, -1.987)$,
and $(a_{RP}, b_{RP}, c_{RP})$ = $(1.199, 1.576, -3.096)$.

We assume that the uncertainties of all observables are Gaussian in nature.
For comparison, Table~\ref{tab:err} lists the expected median uncertainties for
a star with $G=17$~mag. This clearly shows the major improvement in \Gaia~DR2,
where even for a much faint star, the astrometric and photometric uncertainties
of \Gaia~DR2 are superior to those of the brighter TGAS sample. For reference,
a star with $G=12.5$ in \Gaia~DR2 is expected to have a median astrometric
uncertainty of 0.02~mas, and its proper motion uncertainty is expected to be
0.03~mas/yr.

This mock catalogue contains $\sim 620$ million stars in total. Following the
error convolution, we again discard stars that have negative parallaxes or
relative parallax uncertainties larger than 30\%. The resulting sample contains
$\sim 200$ million sources of which only $\sim 0.25\%$ belong to the
halo component. Note that, even though this cut decreases the total
number of stars by a factor of 3, it removes 99\% of the halo stars.

The \Gaia~DR2 catalogue will contain full phase-space information only for stars
brighter than $G \sim 12.5$. Thus in the test cases in which we use stars that
have full phase-space information, we only consider stars that have magnitudes
brighter than $G=12.5$. Even though the second data release of \Gaia\, will not
feature metallicity estimates, in our first test case we do use \feh\, together
with the astrometric, velocity and photometric data during the training and
evaluation in order to gauge the optimal performance of the classifier. In the
second scenario, we use the features we know are going be available in the
actual data release in this bright magnitude range. These first two scenarios
are equivalent to those when we used the GUMS sample in the Solar neighbourhood
convolved with TGAS errors, but now the GUMS data is convolved with the expected
errors for \Gaia~DR2.

In the 3rd test case, we use our entire \Gaia~DR2-like mock catalogue to
identify potential halo stars, using only the 5D astrometric solution and the
photometric data. Since the stars in this entire mock catalogue are no longer
concentrated in the Solar neighbourhood but span a much larger volume, we
convert their positions and velocities to a Galactic cylindrical coordinate
system $(R, \phi, z, \tilde{v}_R, \tilde{v}_{\phi}, \tilde{v}_z)$. For the
velocity transformations, we assume the line-of-sight velocity of each star to
be 0~\kms. From the 200 million sources that comprise this mock catalogue, we
use 150 million for the training process, and the rest to evaluate the model.
During the splitting of the data, we made sure there are equal proportions of
stars in a given magnitude range to the total number of stars in both the
training and the test set.

The performance of the classifier for each of the three scenarios outlined above
is listed in Table~\ref{tab:GUMS}. In the cases when full phase-space
information is available, the classifier recovers at least 90\% of the halo
stars in the test set, with less than 2\% contamination. In fact, its
performance is nearly equally as good as when trained and applied on the ideal,
error-less data. This is mainly due to the high precision measurements expected
for the \Gaia~DR2 data.

\begin{figure*}
\centering
\includegraphics[width=\textwidth]{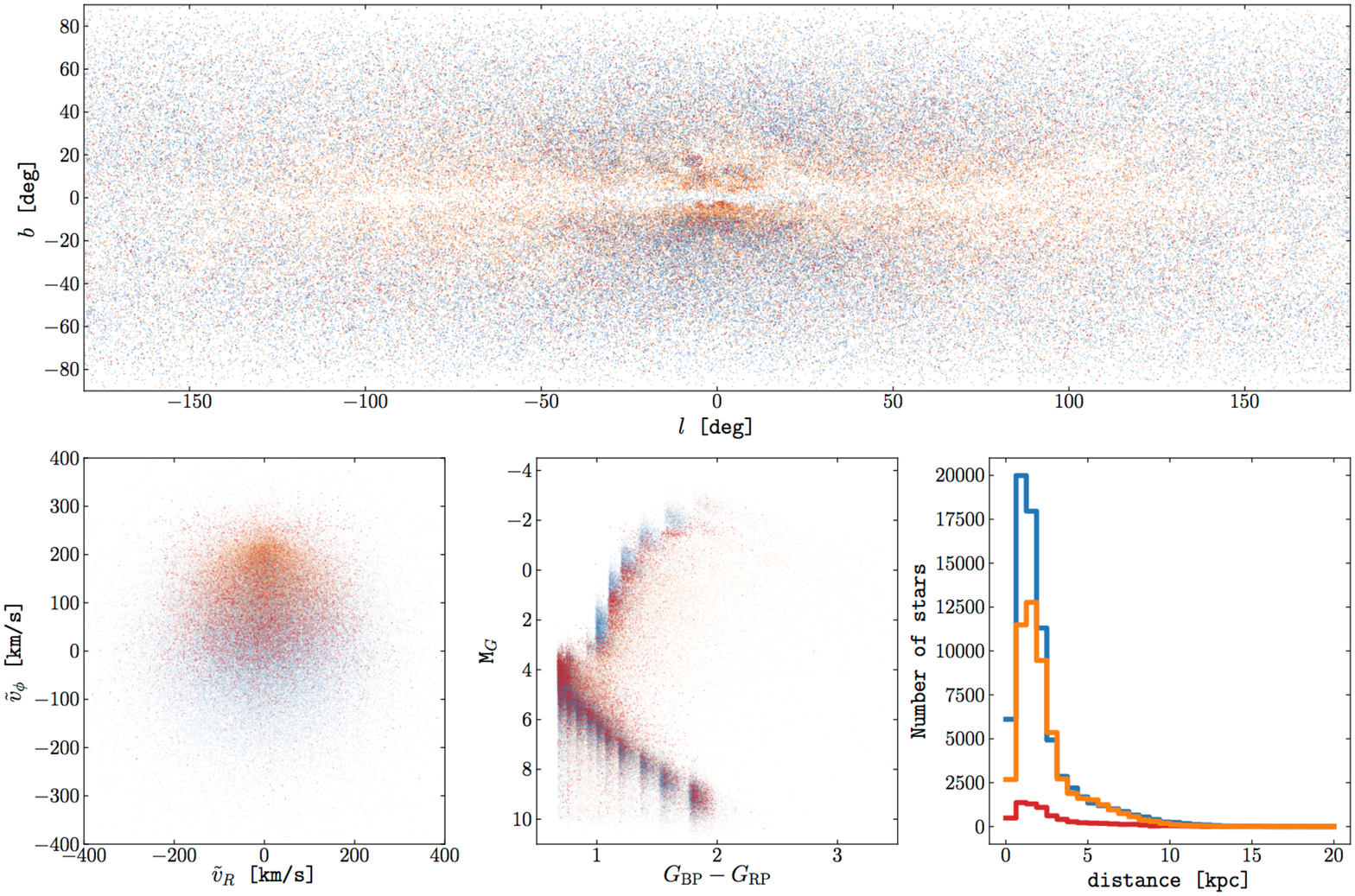}
\caption{The performance of the classifier when applied to the error convolved
GUMS data with \Gaia~DR2 uncertainties, in the case when only the 5D phase-space
coordinates and the optical photometry of the stars are used as training
features. Top panel: the sky distributions of the correctly detected
halo stars (blue), false negatives (orange), false positives (red). Bottom left:
the velocity distribution in the Galactic cylindrical pseudo velocity plane
$\hat{v}_{\phi}$ vs $\hat{v}_R$ (\vlos = 0~\kms). Bottom
middle: a HR diagram of the stars shown on the top panel. Symbols are as on the
top panel. Bottom right: the distance distribution of the stars in the sample.
The histogram colours match the colours of the points in the other panels.}
\label{fig:dr2-5d-full-test}
\end{figure*}

In the final test case, in which we do not know the line-of-sight velocities of
the stars, our classifier recovers nearly 60\% of the stars in the test set,
with less than 10\% contamination. The top panel on
Figure~\ref{fig:dr2-5d-full-test} shows the on-sky distribution of the correctly
classified halo stars (blue points), the false positives (red points) and the
false negatives (orange points) for this case. One can see that the stars
correctly classified as halo are uniformly distributed on the sky, while the
stars wrongly labelled as halo or non-halo, tend to inhabit the regions
closer to the Galactic plane ($|b|<25$~deg).

The bottom left panel on the same Figure shows the velocity distribution of the
stars in the Galactocentric cylindrical coordinate system. One can see that
the majority of the stars wrongly classified as non-halo are centred at
$\tilde{v}_{\phi} \sim 180$~\kms, and have kinematics similar to those of disk
stars. The bottom middle panel shows a HR diagram where the
wrongly labelled non-halo stars are seen to have systematically redder colour,
or to be systematically fainter than the correctly recovered halo stars. From
this we conclude that our classifier only has problems to identify halo stars
that have kinematics similar to those of disk stars, while at the same having
redder \Gbp-\Grp\, colours.

The bottom right panel in Figure~\ref{fig:dr2-5d-full-test} shows distance
distributions from the Sun for the stars correctly classified as halo
(blue histogram), stars wrongly classified as non-halo (orange histogram), and
stars falsely labelled as halo (red histogram).

\begin{figure*}
\centering
\includegraphics[width=\textwidth]{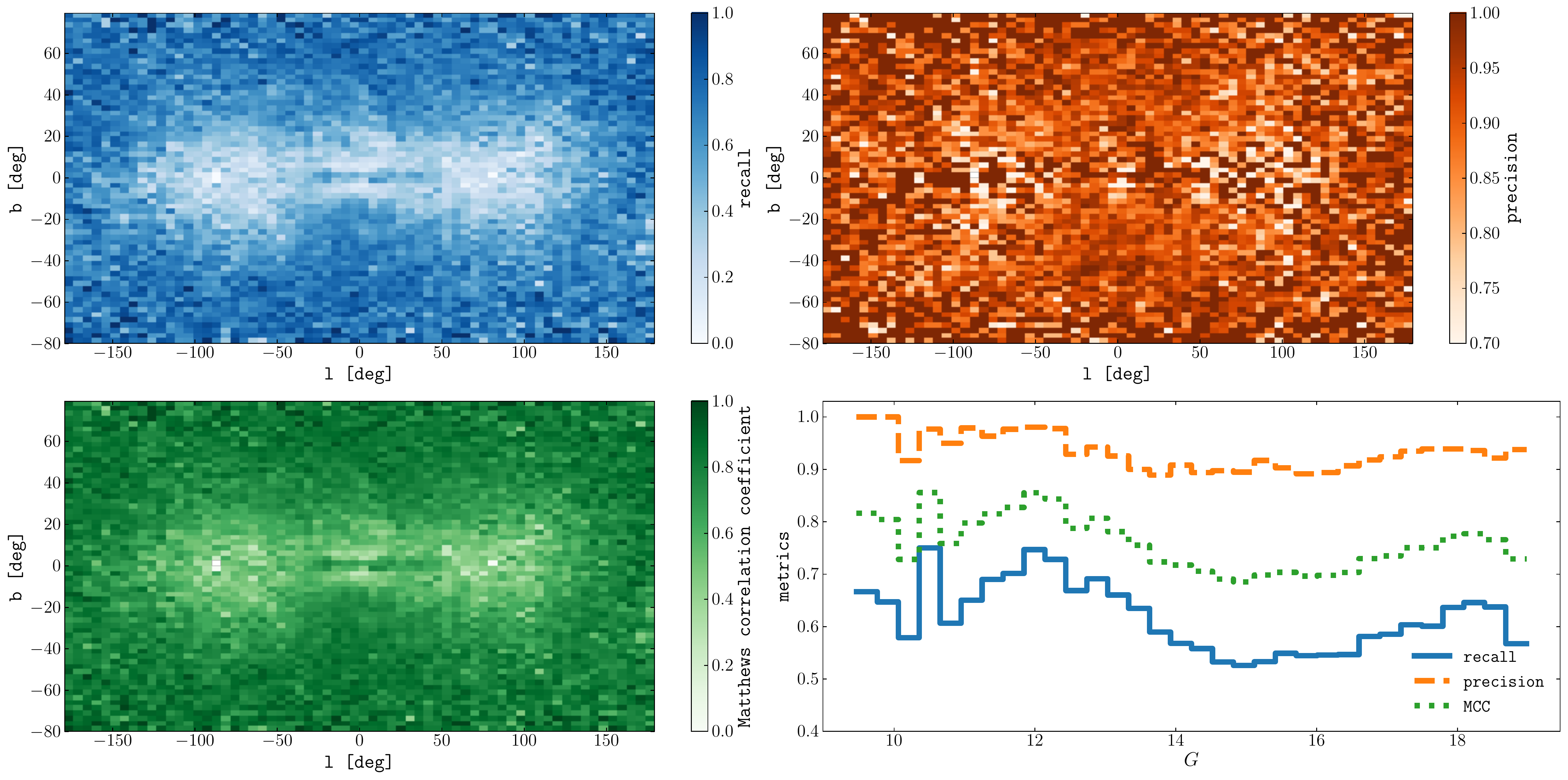}
\caption{The performance of the classifier as a function of Galactic
$(l,b)$ coordinates and the apparent magnitude $G$ for each of the three key
diagnostic metrics, when applied \Gaia~DR2 like data (without line-of-sight
velocity information, i.e. the dataset
used also in Fig~\ref{fig:dr2-5d-full-test}.}
\label{fig:dr2-5d-full-evaluate}
\end{figure*}

The top and bottom-left panels in Figure~\ref{fig:dr2-5d-full-evaluate}
display the performance of our classifier as a function of Galactic $(l,b)$
coordinates according to the three key diagnostic metrics (recall, precision
and the Mathews correlation coefficient) we use to evaluate our final test case.
One can see that the model detects halo stars rather well across the sky, with
the exception of the region spanned by the Galactic disk where we see an
expected drop in performance, particularly in the recall and the Mathews
correlation coefficient. On the other hand, the precision of the model is rather
uniform across the entire sky, meaning that our halo sample will have negligible
level of contamination even in the direction of the disk. In the bottom right
panel of the same Figure we show how the classifier performs as a function of G
magnitude. The panel shows that the model is most effective for the bright
nearby sample of stars. Note that the precision again remains exceptionally
high for any magnitude bin.

\begin{table}
\centering
\caption{Median standard deviations of the uncertainty distributions of the
observables with which we convolve the ideal GUMS selected data to create more
realistic catalogues. In the case of \Gaia~DR2, the uncertainties are those
for a star with $G = 17$.}
\label{tab:err}
\begin{tabular}{lccc}
\hline
\hline
Observable      & $\sigma_{\rm TGAS}$ & $\sigma_{\rm DR2}$($G=17$) & unit     \\
\hline
RA              & 0.23                & 0.10                & [mas]    \\
Dec             & 0.21                & 0.09                & [mas]    \\
parallax        & 0.32                & 0.12                & [mas]    \\
$\mu_{\rm RA}$  & 0.99                & 0.19                & [mas/yr] \\
$\mu_{\rm Dec}$ & 0.83                & 0.17                & [mas/yr] \\
\vlos           & 1.08                & 2.50                & [\kms]   \\
$G$             & $3\times10^{-2}$    & $9\times10^{-3}$    & [mag]    \\
\Gbp            & $3\times10^{-2}$    & $8\times10^{-2}$    & [mag]    \\
\Grp            & $3\times10^{-2}$    & $7\times10^{-2}$    & [mag]    \\
\feh            & 0.25                & 0.25                & [dex]    \\
\hline
\end{tabular}
\end{table}

\subsection{Comparison to the classical $kNN$ algorithm}
\label{subsec:algcomp}

As an additional test on the suitability of our boosted trees model to identify
halo stars, it is useful to compare its performance to other classification
algorithms or methods. In this section, we compare the
performance of our model to that given by the classical $k$-Nearest-Neighbours
($kNN$) algorithm.

The $kNN$ \citep[e.g.][]{Cover:2006:NNP:2263261.2267456} is one of the classical
machine learning algorithms for classification and pattern recognition. In its
most basic form the algorithm is quite simple. The training phase only stores
the feature vectors and the classification labels of the training data. A new
data point is then labelled to have the most frequent class of its nearest $k$
neighbours (hence the name) in the space spanned by the feature vectors being
considered. It is quite common to use the Euclidean distance. One could extend
the algorithm by weighting the distances between the data we are trying to
predict and its $k$ nearest neighbours, give more weight to certain features, or
consider different distance metrics for example. Here, we use the $kNN$ method
in it is simplest form, as implemented in the \texttt{scikit-learn} library.

We proceed to apply the $kNN$ algorithm to our mock data with \Gaia~DR2
uncertainties, and we use the same training features subject to the same
quality cuts as in the case of the boosted trees model. Prior to applying the
algorithm, we scale each feature of the training and test data to have zero mean
and unit variance, which is standard  procedure when using the $kNN$ algorithm
with a Euclidean distance metric. Then we run the algorithm through our training
dataset with different values for $k$,  the number of neighbours to be considered for
the classification process. We found $k=31$ to give the optimal performance.
Applying this method on the bright mock sample with \Gaia~DR2 errors, for which we have
full phase space information, the $kNN$ model achieves a recall of 0.59 and a
precision of 0.96. When applied on the full \Gaia~DR2 mock sample, with 5D phase
space information, the $kNN$ classifier research a recall of 0.42 and a
precision of 0.92. For this test, we see that, while the $kNN$ performs
admirably well, the boosted trees model is superior in detecting a larger
fraction of halo stars. This is not surprising, as the $kNN$ method is typically
used as a clustering algorithm, while the halo stars are not clustered in the
feature space we considered. In addition, $kNN$ does not scale well for large
datasets such as our \Gaia~DR2 sample, since in order to assign a class label to
each new data point, it needs to calculate the distance to all samples in the
training set. For the dataset used in this work, the $kNN$ algorithm is
$\sim~4$ times slower compared to the \xgb\, model.

\section{Detecting halo stars in TGAS}
\label{sec:tgas}
\begin{figure*}
\centering
\includegraphics[width=\textwidth]{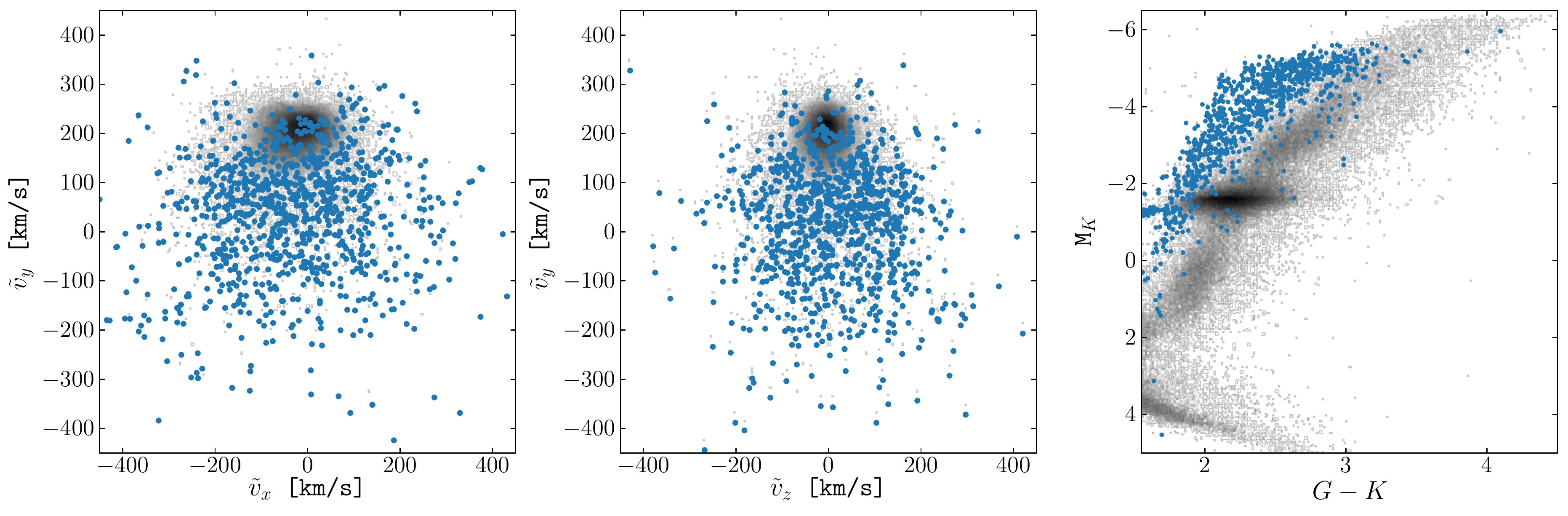}
\caption{The TGAS$\times$RAVE data used to train the TGAS halo star classifier.
The blue points mark the locations of high confidence halo stars selected on the
bases of their metallicities and kinematics, in an entirely model independent
way. The underlying density map shows the remaining stars after imposing our
quality criteria on a logarithmic scale (see Section~\ref{subsec:training} and
\citet{2017A&A...598A..58H} for details).}
\label{fig:tgas_train_set}
\end{figure*}

The exercises we described in the Section~\ref{sec:gums} are entirely model
dependent: the definition of what constitutes a halo star is completely defined
by the GUMS data model. They are meant as a proof of concept, aiming to show the
capabilities of our halo classifier given the confines of the data model.

In this Section, we apply our halo classifier to the entire TGAS dataset. In
what follows we describe the creation of our training set, and how we assign
the halo labels. Later in Section~\ref{subsec:tgas_halo} we apply the trained
classifier to the TGAS dataset. The approach we present here is fully model
independent, in the sense that we use an entirely data driven approach to define
the halo labels using an extended number of stellar features, and then train
the classifier to be able to detect halo stars based on a limited feature set.

\subsection{Defining the training sample}
\label{subsec:tgas_train}

In order to detect halo stars in an entirely model independent manner, we follow
the example set by \citet{2017A&A...598A..58H} which combined metallicity and
kinematics criteria exploiting the synergy between TGAS and RAVE. For a more
detailed discussion on this selection process we refer the interested reader to
Section~2 in \citet{2017A&A...598A..58H}. Here we provide just a summary of the
selection criteria and steps to label stars as halo. In this work we use the
TGAS dataset cross-matched to RAVE from \citet{2017arXiv170704554M} which used
the TGAS parallaxes as priors to derive the spectrophotometric parallaxes.

We only consider stars that have a velocity uncertainty $\Delta_{RV}\le10$~\kms,
{\sf CorrCoeff} $\ge 10$, SNR $\ge 20$ and {\sf algoConv} $\ne 1$. These
criteria ensure that the stars in the RAVE data have reliable radial velocities
and astrophysical parameters. In addition to this, we again only select stars
that have relative parallaxes uncertainty better than 30\%. We use the updated
spectrophotometric parallaxes from RAVE whenever they have better relative
uncertainties compared to the parallaxes from TGAS.

In order to label halo stars after imposing the above quality selection
criteria, similarly as in \citet{2017A&A...598A..58H}, we select stars that have
[M/H] $\le -1$ and distances greater than 0.1~kpc. In this work we use the
metallicity estimates provided by \citet{2017arXiv170704554M} which are
calibrated to match external catalogues, and this is why the metallicity
threshold is different here compared to \citet{2017A&A...598A..58H}.

The improved \logg, \Teff, and [M/H] by \citet{2017arXiv170704554M} leads to a
low contamination level by disk stars in the metal-poor halo candidates as
judged by looking at their velocities. Still, some disk contamination remains,
and we model this by fitting a two component Gaussian Mixture Model to the
Cartesian velocity coordinates $(v_x, v_y, v_z)$. One of the Gaussians is
centred at $v_y \sim 0$~\kms, and the stars that have the higher probability
to be drawn from this component are labelled to be halo stars in our training
set. The other component is centred at on $v_y \sim 180$~\kms, and the stars
that have a higher probability to be drawn from this Gaussian are eliminated
as disk contaminants. The mean velocity of the disk contaminants in this case
seems to be lower than typically assumed velocity of the Local Standard of Rest,
most likely because our low metallicity sample is mostly contaminated by
thick disk stars. With this approach we label 1217 stars in total to be halo.

Figure~\ref{fig:tgas_train_set} shows the training set based on the cross-match
between TGAS and RAVE. The first two panels show the location of the halo stars
(blue points) in the Cartesian velocity space. Although the Gaussian velocity
decomposition was done in $(v_x, v_y, v_z)$, here we show
$(\tilde{v}_x, \tilde{v}_y, \tilde{v}_z)$ since these features are used in the
training process. The rightmost panel shows the location of the halo stars in a
HR diagram, for which we used 2MASS photometry
\citep{2006AJ....131.1163S}. Note that the entire TGAS dataset overlaps with the
2MASS catalogue, and since \Gaia~DR1 released photometry only in the broad $G$
band, we rely on the 2MASS photometry for the construction of HR
diagrams when dealing with the TGAS data.

Since nearly all of the halo stars identified in the TGAS~$\times$~RAVE data are
located on the red giant branch, there are not enough main sequence halo stars
to reliably fit the model. Thus, to lower contamination, we introduce a
$G-K > 1.55$ colour cut in the training set. Finally, our complete
TGAS~$\times$~RAVE training set comprises 119728 stars, of which 999 are
labelled as halo.

\begin{figure*}
\centering
\includegraphics[width=\textwidth]{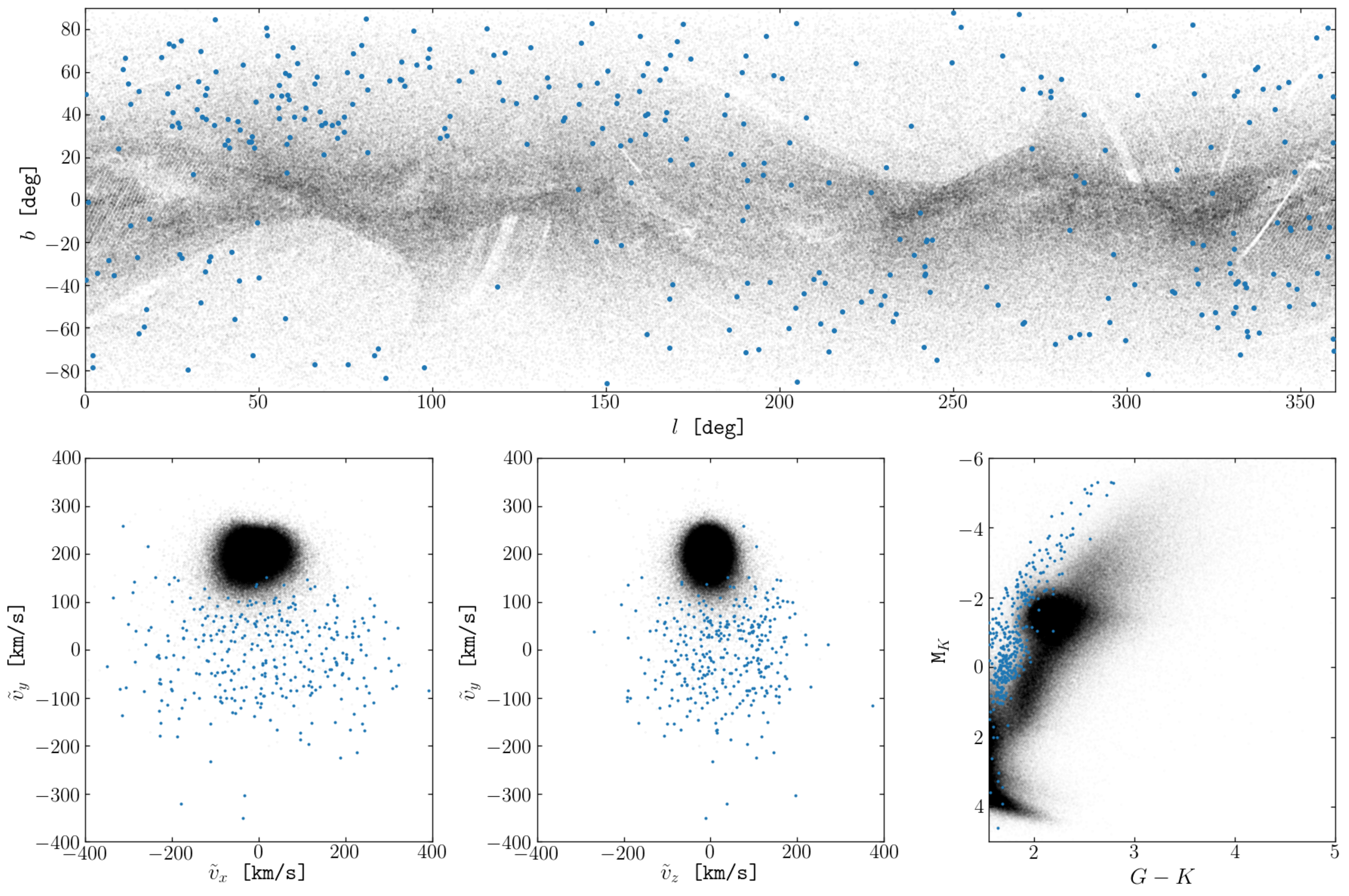}
\caption{On-sky positions, velocity distributions, and a HR
diagrams of the TGAS data fed to the halo classifier. The stars have relative
parallax uncertainties better than 30\% and distances greater than 0.1~kpc.
The 337 blue points mark the locations of the halo candidates, while the black
dots are rest of the stars. One can see that the halo candidates have kinematics
consistent with them being halo, and are located in regions of the
HR typically associated with metal-poor giants.}
\label{fig:tgas-class}
\end{figure*}

\subsection{Classifying halo stars in TGAS}
\label{subsec:tgas_halo}

We train our halo star classifier on the TGAS~$\times$~RAVE training set
following the same procedure as described in Section~\ref{subsec:training}.
For training the model we use the positions of the stars ($x$, $y$, $z$), the
velocities $(\tilde{v}_x, \tilde{v}_y, \tilde{v}_z)$, the $G-K$ colour and
the absolute magnitude M$_{K}$. After optimizing and training the classifier,
we obtain a precision of $0.90 \pm 0.03$ and a recall of $0.63 \pm 0.04$ on a
fully unseen test sample, which comprises 30\% of the complete training set.
The statistics from this training process are better than those for the GUMS
sample of stars selected in the Solar neighbourhood convolved with TGAS
uncertainties. This is because in this case, we are using the updated
spectrophotometric parallaxes from \citet{2017arXiv170704554M}, which have
smaller uncertainties compared to the TGAS parallaxes.

Prior to feeding the TGAS dataset into the halo classifier, we apply certain
selection criteria assuring that we are using appropriately high quality data.
Similarly as when constructing the training dataset, we eliminate all stars in
TGAS that have relative parallax uncertainties larger than $30\%$. In addition,
to lower contamination from nearby disk dwarfs, we exclude all stars with
distances smaller than 0.1~kpc. This leaves us with 423431 stars.

Feeding this dataset into our halo classifier, we detect 337 highly probable
halo candidate stars. Figure~\ref{fig:tgas-class} shows the on-sky and
velocity distributions of the halo star candidates (blue points), as well
as their location on a HR diagram. From this Figure, one can see
that all stars have kinematics consistent with being halo, and are located in
the region of the HR magnitude diagram typically associated with
metal-poor stars. From the top panel on Figure~\ref{fig:tgas-class}, we see
that the identified halo stars are not distributed uniformly on the sky. This is
mainly due to the training set not covering the entire sky. In addition, since
the training set is partially a sub-sample of the input TGAS data, we find 113
halo stars in common between the supervised classification of TGAS, and the
unsupervised classification based on TGAS~$\times$~RAVE.

At first glance, detecting only 337 halo stars may seem too small a number given
the input catalogue of 423431 sources, or may indicate a poor performance of the
classifier. The small number of detections seems to be directly related to the
parallax uncertainties in the TGAS dataset. When constructing our training
sample using TGAS~$\times$~RAVE, the vast majority of the labelled halo stars
are more distant giants for which we used the superior spectrophotometric
parallaxes from RAVE \citep{2017arXiv170704554M}. In fact, only 61 out of the
1217 halo stars we detected in TGAS $\times$ RAVE have parallaxes coming from
the TGAS dataset. Of those, only 16 are giant stars, while the remaining 45 are
closer dwarf stars. Furthermore, the fraction of halo stars we detect in TGAS
(0.079\%) is not very different from the fraction of halo stars present in our
TGAS-like sample drawn from GUMS ($\sim 0.046$\%). In addition, the near
absence of dwarf halo stars in the training set makes our model heavily biased
towards the detecting of red giant halo stars.

In order to asses whether the classifier is performing sensibly, we looked at
the very basic kinematic properties of the selected halo sample. The mean
velocity of the halo component in Cartesian coordinates is
$(\tilde{v}_x, \tilde{v}_y, \tilde{v}_z) = (14, -10, 28)$~\kms, while their
associated velocity dispersions are $(156, 90, 99)$~\kms, respectively. These
values are broadly consistent with the mean velocities
$(\tilde{v}_x, \tilde{v}_y, \tilde{v}_z) = (-32, -34, 13)$~\kms, and their
associated velocity dispersion $(152, 130, 119)$~\kms of the halo sample in the
TGAS~$\times$~RAVE training set. The differences are likely due to the
different number of stars, as well as the different sky coverage and radial
range probed by the different samples. The halo sample selected from
TGAS~$\times$~RAVE data extends out to nearly 7~kpc and covers almost half of
the sky, while the sample of halo stars identified from the TGAS data with our
classifier extends only out to approx 1.3~kpc but covers the entire sky.

\begin{figure}
\centering
\includegraphics[width=8cm]{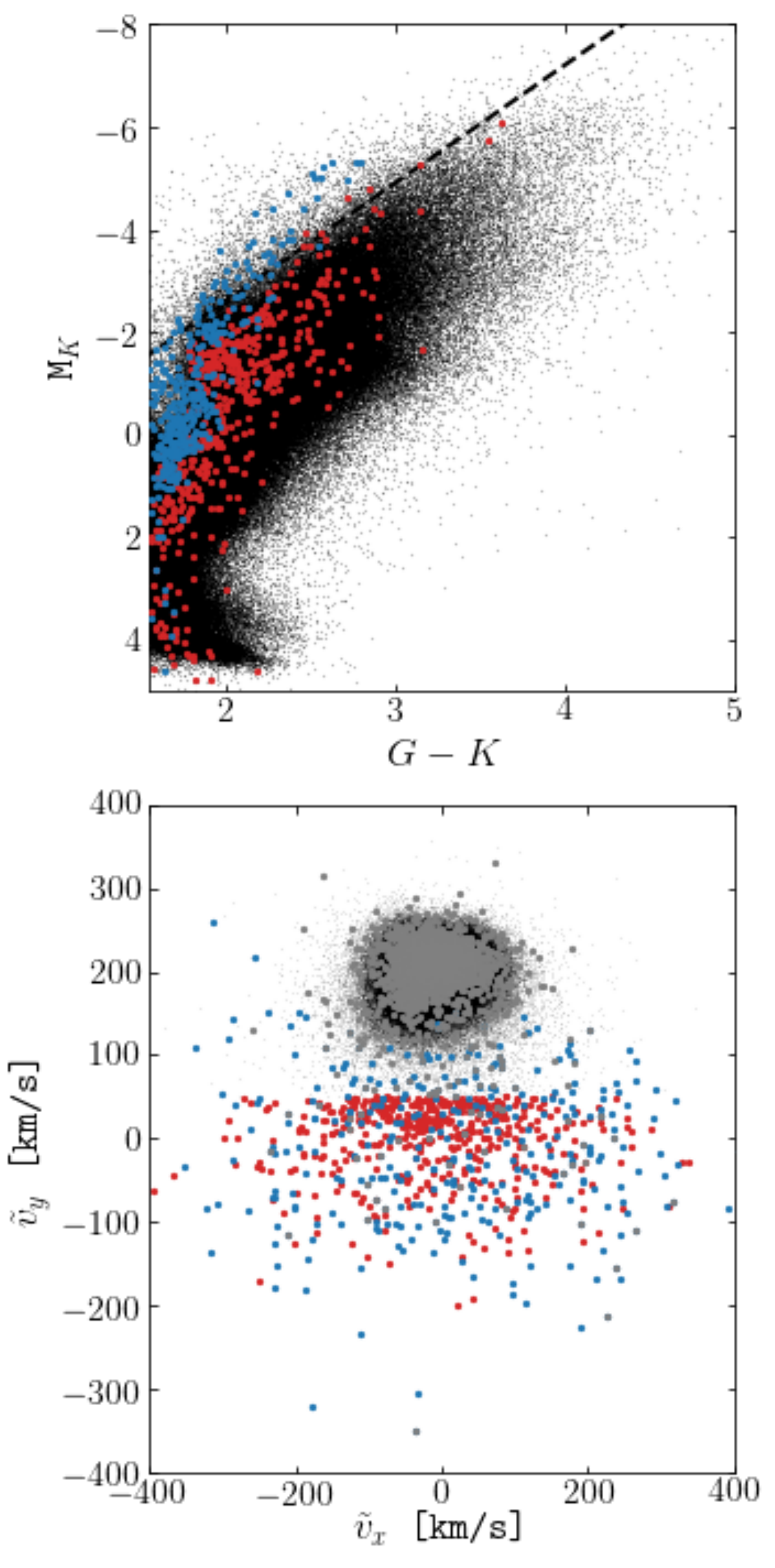}
\caption{The red points are TGAS stars that have $\tilde{v}_y < 50$~\kms and are
not tagged as halo by our classifier, despite having halo kinematics. Those
stars are found to have redder $G-K$ colours than expected for metal-poor halo
stars. It is possible that these stars belong to the metal-poor tail of the
thick disk. The grey points (bottom panel) have M$_K$ magnitudes brighter than
the cut marked by the black dashed line (top panel) and are not labelled as halo
by our classifier. Even though they have colours consistent with those of metal
poor stars, the kinematics of the vast majority of these sources is purely
consistent with that of disk stars.}
\label{fig:tgas-sanity}
\end{figure}

To further test the reliability of the classifier, we do the following sanity
checks. Shown with red points in Figure~\ref{fig:tgas-sanity}, there are 465
stars with $\tilde{v}_y < 50$~\kms\, that are not selected by our classifier
despite having kinematics consistent with halo stars. We find that these stars
have redder $G-K$ colours compared to the halo stars in the training set, and
stars selected to be halo by the classifier. It is possible that some of these
may be in regions of high extinction which makes their colours redder than
expected. The other possibility is the existence of a halo population that is
more metal rich than our unsupervised, metallicity based selection of halo stars
in the training set \citep{Bonaca2017}.

We also notice that there is a considerable number of stars not classified as
halo having brighter M$_K$ magnitudes than the dashed line on the top panel in
Figure~\ref{fig:tgas-sanity}. This region of the HR diagram is
heavily populated by halo stars in the training set. If we look at their
velocity distribution, we find them to have purely disk kinematics. Large
parallax uncertainties are one likely cause for them moving to the bright region
of the HR diagram. When obtaining the distance to a star by inverting a parallax
with a large uncertainty, one may overestimate the distance since its
probability distribution function has an extended tail towards large distances
\citep[see Appendix A in][]{2017A&A...598A..58H}. Overestimating the distance
leads to an overestimation of the absolute magnitude of a star, which may
explain the location of these disk stars on the HR diagrams in
Figures~\ref{fig:tgas-class}~and~\ref{fig:tgas-sanity}.

It might also be useful to compare the results of our model to
the recent study by \citet{2017arXiv171104766P}, who showed that one
can identify halo stars with a physically motivated model, in which
each Galactic component is described by a distribution function. One
can also apply such a model in cases in which the line-of-sight
velocity of the stars is not known by marginalizing over the missing
observable.

When applying the dynamical model of
\citet{2017arXiv171104766P} to the TGAS dataset, we identify 1667
halo candidate stars. This larger number of candidates in comparison
to our technique, is likely driven by the dynamical model not using
metallicity information and not being biased against the detection
of halo dwarfs as in the case of our classifier, although it is
affected by assumptions on the Milky Way's gravitational potential
and the spatial and velocity distribution of each Galactic
component. In an attempt to make a ``fairer'' comparison between the
two procedures of identifying halo stars, we count the number of
candidates identified by the dynamical model that satisfy $1.55\le
(G-K) < 2$ and $M_K < 2$. These cuts essentially select the metal
poor giant stars, since our classifier has been trained to detect
metal poor halo stars, and is practically insensitive to the
existence of halo dwarfs due to the limitations of the training
set. We find that 265~halo star candidates have been identified by
the dynamical model that satisfy these conditions, a number very
similar to the 277~candidates identified by our gradient boosted
tree classifier given the same cuts.

\section{Conclusions}
\label{sec:concl}

The stellar halo is an essential component needed to understanding the
assembly process and some of the key properties of our Galaxy. A sizeable sample
of halo stars can help us to constrain the Milky Way's mass or the shape of its
gravitational potential. By searching for phase space substructure in the stellar
halo we may be able to isolate some of the primordial building blocks of the Galaxy.

In this contribution we trained a series of gradient boosted trees machine
learning models and assessed their ability to identify halo stars in the
publicly accessible \Gaia~DR1 and the upcoming \Gaia~DR2 catalogues, based on
the available astrometric and photometric data.

We first tested the performance of our models on a data sample selected from the
\Gaia\, Universe Model Snapshot (GUMS) with a TGAS-like selection function. When
using the full phase-space information and the optical \Gaia\, photometry, the
model identifies over 90\% of the halo stars in an entirely unseen dataset.
When the training data lacks \vlos\, information, the model recovers
over 85\% of the halo stars in an unseen test set. In both of these ideal cases,
the level of contamination in the labelled halo sample is negligible
($<1\%$).

To investigate the performance of our halo star classifier in more realistic
cases, we convolved a large GUMS sample with stars having $6 < G <19$ with
uncertainties expected for \Gaia~DR2. For stars with relative parallax errors
$< 30\%$, the model is able to identify $\sim 60\%$ of the halo stars in the
above magnitude range when using \Gaia\, photometry and 5D phase-space
information only as training features. For magnitudes brighter than
$G \approx 12.5$, full astrometric solutions should be available in \Gaia~DR2,
and coupling that with the optical photometry from the satellite we expect that
it will be possible to correctly identify at least $\sim 90\%$ of the halo stars
that have reasonable distances. From the above experiments, we conclude that
even though lack of \vlos\, does decrease the performance of our classifier, we
are still able reliably identify a large fraction of halo stars, with negligible
level of contamination.

It is worth noting that the our requirement that stars have a relative parallax
uncertainty $< 30\%$ considerably reduces the reach of our method. In our
\Gaia~DR2-like sample selected from GUMS, 90\% of the halo stars that satisfy
this criterion are within 5~kpc from the Sun, and this constitutes only
$\sim 2\%$ of the full halo population. Efforts that aim to improve parallax
uncertainties \citep[e.g.][]{2017arXiv170605055A} may therefore be crucial and could
lead to a much higher fraction of the halo being accessible. Engineering
informative training features that do not require a direct use of the distance
but could employ the parallax directly without introducing significant biases
could in principle extend this method to samples spanning larger volumes.

In order to apply our halo classifier on the TGAS data, we trained it on a
sample of halo stars identified in the TGAS~$\times$~RAVE dataset published by
\citet{2017arXiv170704554M}, which contains updated spectrophotometric
parallaxes that use the TGAS parallaxes as priors. The halo sample in the
TGAS~$\times$~RAVE was selected via a metallicity cut and it was further cleaned
by unsupervised kinematic modelling, in an entirely data driven way. Here
we focused on the stars that belong to the red giant branch since very few
halo stars in the TGAS~$\times$~RAVE data belong to the main sequence, and they
are insufficient to properly train the model in that regime.

We apply the classifier trained on the TGAS~$\times$~RAVE data on the entire
TGAS catalogue, and identify 337 high confidence red giant branch halo stars.
While this number may seem very small at first glance, it is because we only
consider TGAS stars that have relative parallax errors smaller than 30\%, which
severely limits the number of red giant branch stars in the sample. Nevertheless
the halo sample selected in this manner has broadly consistent kinematics with
the halo sample selected from the TGAS~$\times$~RAVE dataset, adding confidence
that our model is performing correctly.

We look forward to applying our halo star classifier on the data coming from the
Second Data Release of the \Gaia\, mission. Our tests show that with this method
we should be able to identify a high confidence halo sample that numbers in the
thousands. Such a sample would be extremely valuable in helping us unravel the
history of our Galaxy.

\begin{acknowledgements}
We would like to thank the anonymous referee for their comments
which improved the manuscript. We thank Anthony Brown for his useful
comments regarding the expected uncertainties in \Gaia~DR2. We are
also very grateful to Yonatan Alexander for his advice regarding the
usage for \xgb\, and \texttt{bayes\_opt}.  This work has been
supported by a VICI grant from the Netherlands Organisation for
Scientific Research, NWO, and by NOVA, the Netherlands Research
School for Astronomy. We have made use of data from the
European Space Agency (ESA) mission {\it Gaia}
(\url{http://www.cosmos.esa.int/gaia}), processed by the {\it Gaia}
Data Processing and Analysis Consortium (DPAC,
\url{http://www.cosmos.esa.int/web/gaia/dpac/consortium}). Funding
for the DPAC has been provided by national institutions, in
particular the institutions participating in the {\it Gaia}
Multilateral Agreement.

In addition to the software discussed in the paper, this work made use of
\texttt{vaex} \citep{2017IAUS..325..299B},
\texttt{numpy} \citep{Walt:2011:NAS:1957373.1957466},
\texttt{matplotlib} \citep{Hunter:2007},
and \texttt{scikit-learn} \citep{scikit-learn}.
\end{acknowledgements}

\bibliographystyle{aa}

\end{document}